# HIGH-FREQUENCY RADIO PROPERTIES OF SOURCES IN THE *FERMI*-LAT 1 YEAR POINT SOURCE CATALOG

Elizabeth K. Mahony[1,2], Elaine M. Sadler[1], Tara Murphy[1,3], Ronald D. Ekers[2], Philip G. Edwards[2], and Marcella Massardi[4]

[1] Sydney Institute for Astronomy, School of Physics, University of Sydney, NSW 2006, Australia
[2] Australia Telescope National Facility, CSIRO, P.O. Box 76, Epping, NSW 1710, Australia
[3] School of Information Technologies, The University of Sydney, NSW 2006, Australia
[4] INAF, Osservatorio Astronomico di Padova, Vicolo dell'Osservatorio 5, I-35122 Padova, Italy

## ABSTRACT

The high-frequency radio sky, like the gamma-ray sky surveyed by the *Fermi* satellite, is dominated by flat spectrum radio quasars and BL Lac objects at bright flux levels. To investigate the relationship between radio and gamma-ray emission in extragalactic sources, we have cross-matched the Australia Telescope 20 GHz survey catalog (AT20G) with the *Fermi*-LAT 1 year Point Source Catalog (1FGL). The 6.0 sr of sky covered by both catalogs ($\delta < 0°$, $|b| > 1°.5$) contains 5890 AT20G radio sources and 604 1FGL gamma-ray sources. The AT20G source positions are accurate to within ∼1 arcsec and, after excluding known Galactic sources, 43% of *Fermi* 1FGL sources have an AT20G source within the 95% *Fermi* confidence ellipse. Monte Carlo tests imply that at least 95% of these matches are genuine associations. Only five gamma-ray sources (1% of the *Fermi* catalog) have more than one AT20G counterpart in the *Fermi* error box. The AT20G matches also generally support the active galactic nucleus (AGN) associations in the First LAT AGN Catalog. We find a trend of increasing gamma-ray flux density with 20 GHz radio flux density. The *Fermi* detection rate of AT20G sources is close to 100% for the brightest 20 GHz sources, decreasing to 20% at 1 Jy, and to roughly 1% at 100 mJy. Eight of the matched AT20G sources have no association listed in 1FGL and are presented here as potential gamma-ray AGNs for the first time. We also identify an alternative AGN counterpart to one 1FGL source. The percentage of *Fermi* sources with AT20G detections decreases toward the Galactic plane, suggesting that the 1FGL catalog contains at least 50 Galactic gamma-ray sources in the southern hemisphere that are yet to be identified.

*Online-only material:* color figures, machine-readable table

## 1. INTRODUCTION

The recent release of the First *Fermi*-LAT (1FGL; Abdo et al. 2010a) catalog provides an unprecedented view of the gamma-ray sky. Many of these high energy sources (particularly those at high Galactic latitudes) are attributed to active galactic nuclei (AGNs), where the gamma-ray emission is thought to originate from collimated jets close to the core (Blandford & Rees 1978). As a result, flat spectrum radio-loud QSOs (FSRQs) and BL Lac objects, where the jet is more closely aligned with our line of sight, dominate the population of extragalactic sources selected in gamma-ray surveys. Further properties of the AGN detected by *Fermi* are presented in more detail in the First LAT AGN catalog (1LAC; Abdo et al. 2010b).

The 1FGL catalog lists AGN associations for 41% of *Fermi* sources with $|b| > 1°.5$, but there are many that remain unidentified. The relatively large uncertainty in the *Fermi* positions (typically 540 arcsec) makes it necessary to search at other wavelengths in order to identify the correct astronomical source, and the *Fermi* team have used a Bayesian analysis method to cross-match the 1FGL with a range of catalogs (Abdo et al. 2010a).

Searching high-frequency radio catalogs provides an alternative and potentially more direct way to identify the extragalactic *Fermi* sources. The dominant emission at high radio frequencies originates from the radio jets close to the central core, and thus provides a snapshot of the most recent AGN activity (Mattox et al. 2001). This is very different from lower radio frequencies near 1.4 GHz, where extended radio lobes, built up over large periods of time, dominate the emission. The source density of high-frequency radio sources on the sky is also lower,[5] greatly reducing the number of spurious radio sources within a *Fermi* error box. Since most *Fermi* sources are expected to have a flat radio spectrum, this makes radio identification with a high-frequency catalog much more straightforward and reliable.

This was recognized by Healey et al. (2008), who assembled an all-sky catalog of 8 GHz measurements of candidate blazars (selected on the basis of their radio spectral index at 1–5 GHz) to help identify *Fermi* sources. The 20 GHz AT20G catalog, which has no spectral pre-selection, provides an independent check of radio–gamma-ray associations for southern sources.

The Australia Telescope 20 GHz (AT20G) survey (Murphy et al. 2010) is a blind radio survey of the southern sky (excluding the Galactic plane at $|b| < 1°.5$) down to a limiting flux density of 40 mJy at 20 GHz. Observations for this survey were carried out using the Australia Telescope Compact Array from 2004 to 2008. This 20 GHz selection produced a catalog dominated by flat spectrum radio sources (Sadler et al. 2006; M. Massardi et al. 2010, in preparation), which is ideal in searching for *Fermi* counterparts. The AT20G radio positions are accurate to ∼1 arcsec, which also allows us to make optical identifications in a reliable way.

---

[5] The surface density of sources stronger than 100 mJy at 20 GHz is 0.15 deg$^{-2}$ (Murphy et al. 2010), a factor of 17 lower than the value 2.65 deg$^{-2}$ for sources stronger than 100 mJy at 843 MHz (Bock et al. 1999).

**Table 1**
Source Statistics for the *Fermi*–AT20G Catalog Comparison

| Region of Sky | *Fermi* Sources |
|---|---|
| Full *Fermi*-LAT catalog (1FGL) | 1451 |
| Southern hemisphere only: $\delta < 0$ and $|b| > 1°\!.5$ | 604 |
| Excluding 16 hr < R.A. < 18 hr and $-15° < \delta < 0$ | 577 |
| Excluding known Galactic objects (1FGL-South data set) | 540 |

**Note.** Since the AT20G catalog only covers the southern hemisphere, our analysis used a subset of the 1FGL catalog covering the same region of sky.

In this paper, we present an analysis of the AT20G sources detected in the 1FGL catalog. Section 2 details the cross-matching procedures between 1FGL and AT20G, while Section 3 presents these results. This is followed by a discussion of the multiwavelength properties, including a comparison between the 20 GHz and gamma-ray fluxes in Section 4. Concluding remarks are presented in Section 5.

## 2. CROSS-MATCHING 1FGL WITH THE AT20G CATALOG

Since the AT20G catalog only covers the southern sky, we used a subset of 1FGL sources south of the equator (and excluding the Galactic plane $|b| < 1°\!.5$; see Table 1) for the cross-matching. An incomplete region in the AT20G survey at R.A. 16–18 hr and $-15° < \delta < 0°$, due to bad weather during the observations, was also excluded. Finally, since the AT20G catalog is best suited for identifying extragalactic AGN, sources classified as Galactic[6] in the 1FGL catalog (including pulsars, pulsar wind nebula, globular clusters, supernova remnants, and X-ray binaries) were removed from the sample. Table 1 gives a breakdown of the numbers left in the 1FGL catalog at each stage in this process. We have used the second version of the 1FGL catalog (updated on 2010 February 4), which was the most recent version at the time of writing.

Using this set of *Fermi* sources (referred to here as the 1FGL-South data set), we then searched the AT20G catalog for radio sources within the 95% confidence ellipse given in the 1FGL catalog. This resulted in 234 (43%) *Fermi* sources with AT20G counterparts. In a few cases, there were two or more AT20G sources within the *Fermi* ellipse, as discussed in Section 3.1, giving a total of 240 AT20G matches to the 540 *Fermi* sources.

To investigate the likely number of spurious matches due to chance, we performed a simple Monte Carlo test. This involved generating a fake 1FGL-South data set, with the same number of sources and 95% confidence parameters, and then searching for AT20G sources. The fake data sets were produced by displacing the *Fermi* positions by 10° in right ascension. This was repeated five times, resulting in an average of 10.6 matches to the 540 *Fermi* sources. Thus, we would expect around 4.4% of the AT20G–*Fermi* matches to be chance associations. By the nature of our cross-matching, we are also missing the 5% of real associations that lie outside the 95% *Fermi* confidence region. We have not searched for these, since our priority is to make a reliable set of matches that provides a "clean" sample. We therefore estimate that our final list of *Fermi*–AT20G matches is 95% complete and at least 95% reliable.

---
[6] We also excluded six sources in the LMC and SMC fields.

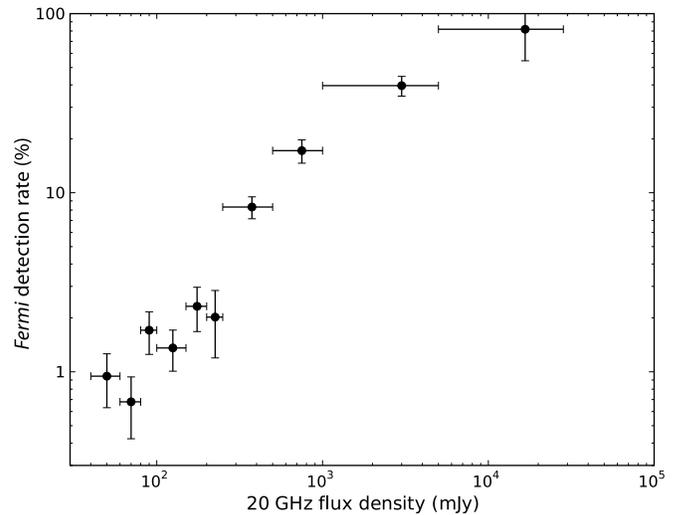

**Figure 1.** *Fermi* detection rate of AT20G radio sources as a function of 20 GHz flux density. The vertical error bars show the counting errors for each bin while the horizontal error bars indicate the bin size.

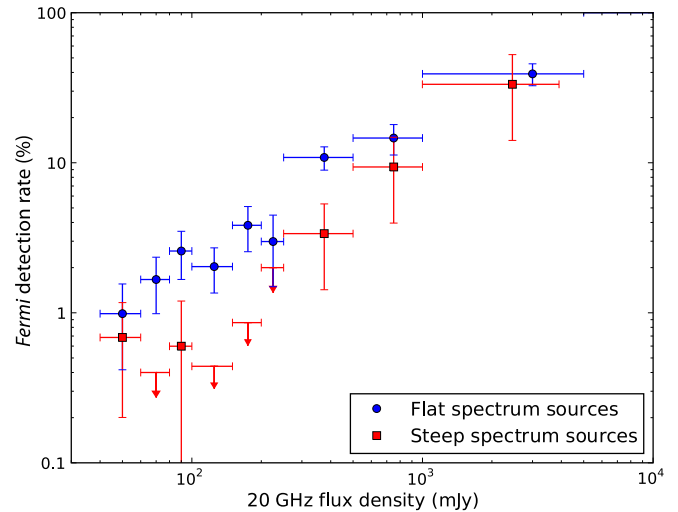

**Figure 2.** *Fermi* detection rate of AT20G radio sources as a function of 20 GHz flux density split into flat and steep spectrum sources. The bins are the same as shown in Figure 1 with the exception of the brightest bin, which has not been plotted as there was only one source above 5 Jy (Centaurus A which has a flat spectrum core). Where there are no detections, an upper limit is plotted indicating the detection rate of one source in that bin.

(A color version of this figure is available in the online journal.)

The *Fermi* detection rate for AT20G sources increases with increasing 20 GHz flux density, as shown in Figure 1. Here, we binned the AT20G catalog in flux density and plotted the percentage of AT20G sources detected in the 1FGL-South data set for each bin plotted. Further correlations with 20 GHz flux density are discussed in Section 4.

Although the AT20G source population is dominated by flat spectrum radio sources, the number of steep spectrum sources increases at lower 20 GHz flux densities (M. Massardi et al. 2010, in preparation). Figure 2 shows the *Fermi* detection rates for the flat and steep spectrum sources, respectively. As expected, the detection rate is much higher for the flat spectrum sources, and there are very few steep spectrum sources detected in gamma rays. Here, we have used the traditional dividing line between the flat and steep spectrum sources of $\alpha = -0.5$ (given by $S \propto \nu^\alpha$) as originally determined by Wall et al. (1977). The physical basis for this definition is discussed further in

**Table 2**
List of Southern *Fermi* Sources That have More than One AT20G Source Inside the *Fermi* 95% Error Ellipse

| *Fermi*-LAT Name | AT20G Association | Offset (deg) | $S_{20}$ (mJy) | $\alpha_5^{20}$ | Listed 1FGL Association |
|---|---|---|---|---|---|
| 1FGL J0141.7−0929 | J014125−092843 | 0.068 | 764 | ... | PKS 0139−09 |
| | J014137−093001* | 0.022 | 99 | ... | ... |
| 1FGL J0625.9−5430 | J062552−543850* | 0.133 | 304 | −0.30 | CGRaBS J0625−5438 |
| | J062648−543214 | 0.132 | 106 | −1.39 | ... |
| 1FGL J0636.1−7521 | J063515−751418 | 0.128 | 47 | ... | ... |
| | J063546−751616* | 0.085 | 3142 | −0.30 | PKS 0637−75 |
| 1FGL J1159.4−2149 | J115910−215005 | 0.074 | 46 | ... | ... |
| | J115921−214246 | 0.113 | 224 | −0.20 | CGRaBS J1159−2142 |
| | J115951−214853* | 0.087 | 308 | −0.71 | PKS 1157−215 |
| 1FGL J1441.7−1538 | J143956−153150* | 0.458 | 569 | ... | PKS 1437−153 |
| | J144145−152336 | 0.244 | 120 | ... | ... |
| 1FGL J2207.1−0021 | J220643−003103* | 0.194 | 477 | ... | CGRaBS J2206−0031 |
| | J220755−000215 | 0.373 | 59 | ... | ... |

**Notes.** The sources that have been used in the following analysis are marked by an asterisk. This is the brightest AT20G counterpart, which also corresponds to the 1FGL AGN association, except in the case of 1FGL J0141.7−0929 where the brightest source falls outside the 95% error ellipse.

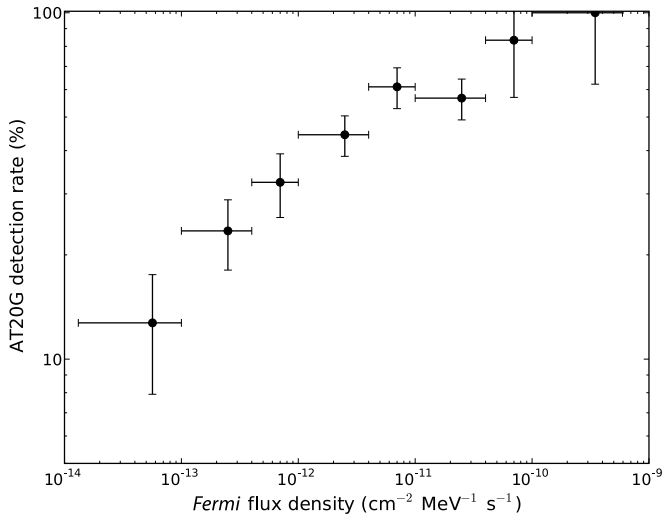

**Figure 3.** AT20G detection rate of the 1FGL-South data set as a function of gamma-ray flux density. The vertical error bars show the counting errors for each bin while the horizontal error bars indicate the bin size.

M. Massardi et al. (2010, in preparation). The spectral index between 8.4 and 20 GHz was used to separate flat and steep spectrum sources. As a result, AT20G sources with no 4.8 or 8.4 GHz information have been excluded in Figure 2. It should also be noted that there are only 12 steep spectrum *Fermi*–AT20G objects that are comparable to the false identification rate of 10.6 sources.

We can also turn this around and look at the AT20G detection rate of the 1FGL-South data set as a function of gamma-ray flux density (Figure 3). The AT20G detection rate also increases with increasing gamma-ray flux density. Here and throughout the paper, we use the gamma-ray flux density given in the 1FGL catalog. This is the differential flux at the pivot energy, which is defined as the energy at which the error on the differential flux is minimal.

A total of 110 of the sources in the 1FGL-South data set are classified as variable in gamma rays, with a less than 1% probability of being a steady source (Abdo et al. 2010a). Of these, 96 (87%) are detected at 20 GHz. On the other hand, only 138 (32%) of the non-variable gamma-ray sources are detected in AT20G. This is discussed further in Section 4.1.

## 3. AT20G SOURCES WITH *FERMI* COUNTERPARTS

### 3.1. Fermi Sources with Multiple AT20G Matches

Table 2 lists the five cases where more than one AT20G source falls within the *Fermi* 95% confidence ellipse. In each case, the brightest 20 GHz source also has a listed association in the 1FGL catalog. These associations have been used in the following analysis. In these cases, it is possible that more than one AT20G source contributes to the gamma-ray flux observed by *Fermi*.

Table 2 also lists the *Fermi* source 1FGL J0141.7−0929, which has two possible AT20G matches but one is outside the *Fermi* 95% confidence ellipse. The *Fermi* team associate 1FGL J0141.7−0929 with the radio source PKS 0139−09 (=AT20G J014125−092843), which has a 20 GHz flux of 764 mJy and a spectral index of $\alpha_1^{20} = 0.06$, but the AT20G position of this source places it just outside the *Fermi* 95% confidence ellipse. The AT20G catalog also lists a second AT20G match (AT20G J014137−093001) with a smaller offset of 0.022 deg. This source is fainter at 20 GHz (99 mJy), but its 1.4 GHz flux density of 150 mJy in the NRAO VLA Sky Survey (NVSS; Condon et al. 1998) implies a flat spectral index of $\alpha_1^{20} = -0.15$. We identify AT20G J014137−093001 as another potential counterpart for the *Fermi* source 1FGL J0141.7−0929, and we use this source rather than PKS 0139−09 in the following analysis.

### 3.2. New Fermi–Radio Associations

Cross-matching the 1FGL catalog with the AT20G catalog revealed a number of new AGN associations, listed in Table 3. These sources do not have any listed counterpart in the 1FGL catalog, though some of them are listed as "affiliations" in the 1LAC (Abdo et al. 2010b). These are classified by the *Fermi* team as candidate matches because the statistical Bayesian analysis, which provides a quantitative association probability, could not be carried out. The fact that these sources appear in the AT20G survey provides further evidence that they are the correct AGN counterparts. The eight sources that do not appear in any of the 1LAC tables are marked by an asterisk in Table 3.

**Table 3**
AT20G Counterparts of *Fermi* Sources with No Listed AGN Association in the 1FGL Catalog

| *Fermi*-LAT Name | AT20G Association | R.A. (J2000) | Decl. (J2000) | Offset (deg) | $l$ | $b$ | $S_{20}$ (mJy) | $\alpha_5^{20}$ | Optical ID |
|---|---|---|---|---|---|---|---|---|---|
| 1FGL J0124.6−0616 | J012450−062501 | 01 24 50.40 | −06 25 01.0 | 0.145 | 145.4 | −67.8 | 51 | ... | Y |
| 1FGL J0315.6−5109 | J031425−510431 | 03 14 25.76 | −51 04 31.7 | 0.218 | 264.3 | −54.0 | 112 | −0.06 | Y |
| 1FGL J0445.2−6008 | J044501−601500 | 04 45 01.60 | −60 15 00.2 | 0.105 | 270.0 | −38.9 | 55 | 0.05 | Y |
| 1FGL J0608.1−0630c | J060746−062307[a] | 06 07 46.16 | −06 23 07.7 | 0.155 | 213.7 | −12.6 | 4157 | ... | N |
| 1FGL J0648.6−6052 | J064740−605805 | 06 47 40.87 | −60 58 05.2 | 0.146 | 270.9 | −23.9 | 79 | 0.01 | N |
| 1FGL J0746.5−0711 | J074627−070951* | 07 46 27.33 | −07 09 51.1 | 0.045 | 225.7 | +8.8 | 93 | ... | N |
| 1FGL J0814.5−1011 | J081411−101210 | 08 14 11.71 | −10 12 10.8 | 0.096 | 231.9 | +13.2 | 98 | ... | Y |
| 1FGL J0904.7−3514 | J090442−351423* | 09 04 42.34 | −35 14 23.8 | 0.018 | 259.4 | +7.8 | 300 | 0.19 | N |
| 1FGL J1029.2−6422 | J103042−641526* | 10 30 42.66 | −64 15 26.2 | 0.200 | 288.4 | −5.4 | 98 | 0.15 | N |
| 1FGL J1153.4−8108 | J115546−810117 | 11 55 46.09 | −81 01 17.7 | 0.154 | 300.7 | −18.4 | 75 | −0.42 | Y |
| 1FGL J1234.0−5736 | J123407−573552* | 12 34 07.04 | −57 35 52.9 | 0.020 | 300.6 | +5.2 | 49 | −0.22 | N |
| 1FGL J1256.1−5922 | J125604−591943* | 12 56 04.93 | −59 19 43.4 | 0.051 | 303.5 | +3.5 | 41 | −0.81 | N |
| 1FGL J1314.9−5338 | J131504−533436* | 13 15 04.24 | −53 34 36.0 | 0.063 | 306.5 | +9.1 | 865 | ... | N |
| 1FGL J1553.5−3116 | J155333−311832 | 15 53 33.49 | −31 18 32.3 | 0.038 | 342.6 | +17.2 | 47 | −0.62 | Y |
| 1FGL J1656.2−3257 | J165616−330207* | 16 56 16.83 | −33 02 07.7 | 0.085 | 350.6 | +6.4 | 287 | 0.19 | N |
| 1FGL J1823.5−3454 | J182338−345412* | 18 23 38.58 | −34 54 12.0 | 0.016 | 358.7 | −9.9 | 84 | −0.24 | N |
| 1FGL J2014.5−0047 | J201428−004723 | 20 14 28.62 | −00 47 23.0 | 0.021 | 42.0 | −18.8 | 49 | ... | Y |
| 1FGL J2022.5−4532 | J202226−451329 | 20 22 26.40 | −45 13 29.2 | 0.311 | 354.9 | −34.5 | 173 | ... | N |

**Notes.** The sources marked with an asterisk are not identified in either of the 1LAC secondary tables.
[a] This source (1FGL J0608.1−0630c) may be a spurious *Fermi* detection and is discussed in Section 3.3.

Many of the new AT20G identifications in Table 3 are at low Galactic latitude ($|b| < 10°$). This is not surprising as the AGN catalogs used by the *Fermi* team to identify the source of the gamma-ray emission are known to be incomplete at low Galactic latitudes (Abdo et al. 2010b).

### 3.3. 1FGL J0608.1−0630c

The AT20G source J060746−062307 (PKS 0605−06) was found within the 95% confidence region of the *Fermi* source 1FGL J0608.1−0630c. This source is in a region of high-mass star formation (known as Mon R2) and was identified by Jackson et al. (2002) as a Galactic H II region. There is also an ultra-compact H II region with associated methanol masers (Walsh et al. 1998), and it is listed as a Galactic source in the AT20G catalog. This source has been flagged as confused in the *Fermi* catalog (denoted by the "c" in the name), meaning that the detection is highly sensitive to the model used to subtract diffuse Galactic gamma-ray emission. As noted in Section 4.7 of the 1FGL catalog paper (Abdo et al. 2010a), this particular source is located close to the Orion nebula, and there is a high chance that it is a spurious *Fermi* detection. If real, 1FGL J0608.10630c is almost certainly Galactic rather than an extragalactic source and has been removed from the following analysis, but the possibility of gamma-ray emission related to extreme mass outflows in this region should be investigated.

### 3.4. The Fermi–AT20G Catalog

The complete list of all 233 *Fermi*–AT20G matches is included as additional material in the electronic version of this paper (see Table 4). The content of each column is as follows.

(1) *Fermi* name (1FGL JHHMM.m-DDMM).
(2) and (3) *Fermi* right ascension and declination.
(4) and (5) Semimajor and semiminor axes of 95% confidence ellipse.
(6) Position angle of 95% confidence ellipse.
(7) Pivot energy in units of MeV. This is the energy at which the error on the differential flux is minimal.
(8) and (9) Gamma-ray flux density and uncertainty in cm$^{-2}$ MeV$^{-1}$ s$^{-1}$. This is the differential flux at the pivot energy.
(10) Gamma-ray variability index. Values greater than 23.21 have <1% probability of being a steady source (Abdo et al. 2010a).
(11) *Fermi* Bright Source Catalogue identification where available (0FGL JHHMM.m-DDMM).
(12) Association as listed in 1FGL.
(13) Class as listed in 1FGL.
(14) AT20G name (AT20GJHHMMSS-DDMMSS).
(15) and (16) AT20G right ascension and declination.
(17) and (18) AT20G Galactic longitude and latitude.
(19) Offset between the *Fermi* and AT20G positions in degrees.
(20)–(22) 20 GHz, 8.4 GHz, and 4.8 GHz flux densities in mJy.
(23) Radio spectral index between 4.8 and 20 GHz.
(24) Quality flag of the AT20G flux. "p" denotes a poor flux measurement for a variety of reasons explained in Murphy et al. (2010). "g" refers to a good flux measurement.
(25) 1 GHz flux density in mJy from either the NVSS or Sydney University Molonglo Sky Survey (SUMSS) catalogs.
(26) and (27) Position of optical identification where available.
(28) $b_J$ magnitude from the SuperCOSMOS database.
(29) and (30) Redshift and the corresponding reference.

### 3.5. Galactic Latitude Distribution of Fermi–AT20G Matches

Figure 4 shows the overall *Fermi* detection rate for extragalactic sources in the AT20G catalog, binned by Galactic latitude. Radio sources identified as Galactic H II regions and planetary nebula in the AT20G catalog were not included in this analysis. Due to the much smaller area of sky at $b > 0$ in the southern hemisphere, the data were binned in $|b|$ rather than $b$ to improve the statistics.

Table 4
The *Fermi*-AT20G Catalog

| *Fermi*-LAT Name | AT20G Name | RA (J2000) | Dec (J2000) | $S_{20}$ | $\alpha$ | z | Ref. |
|---|---|---|---|---|---|---|---|
| 1FGL J0000.9-0745 | AT20GJ000118-074626 | 00:01:18.04 | −07:46:26.8 | 177 | ... | ... | ... |
| 1FGL J0004.7-4737 | AT20GJ000435-473619 | 00:04:35.65 | −47:36:19.0 | 868 | −0.03 | 0.875 | E. K. Mahony 2010, in prep. |
| 1FGL J0013.1-3952 | AT20GJ001259-395426 | 00:12:59.89 | −39:54:26.4 | 1609 | −0.16 | ... | ... |
| 1FGL J0017.4-0510 | AT20GJ001735-051241 | 00:17:35.84 | −05:12:41.5 | 186 | ... | 0.227 | 1LAC |
| 1FGL J0029.9-4221 | AT20GJ003017-422446 | 00:30:17.51 | −42:24:46.3 | 478 | −0.05 | 0.495 | Jackson et al. 2002 |
| 1FGL J0038.4-2504 | AT20GJ003814-245901 | 00:38:14.72 | −24:59:01.9 | 1125 | 0.46 | 0.498 | Jones et al. 2009 |
| 1FGL J0047.3-2512 | AT20GJ004733-251717 | 00:47:33.08 | −25:17:17.7 | 608 | −0.90 | 0.001 | 1LAC |
| 1FGL J0048.0-8412 | AT20GJ004426-842241 | 00:44:26.26 | −84:22:41.8 | 484 | −0.16 | 1.032 | 1LAC |
| 1FGL J0049.8-5738 | AT20GJ004959-573827 | 00:49:59.48 | −57:38:27.6 | 1872 | −0.05 | 1.797 | Peterson et al. 1976 |
| 1FGL J0050.0-0446 | AT20GJ005021-045221 | 00:50:21.59 | −04:52:21.0 | 279 | ... | 0.920 | 1LAC |

(This table is available in its entirety in a machine-readable form in the online journal. A portion is shown here for guidance regarding its form and content.)

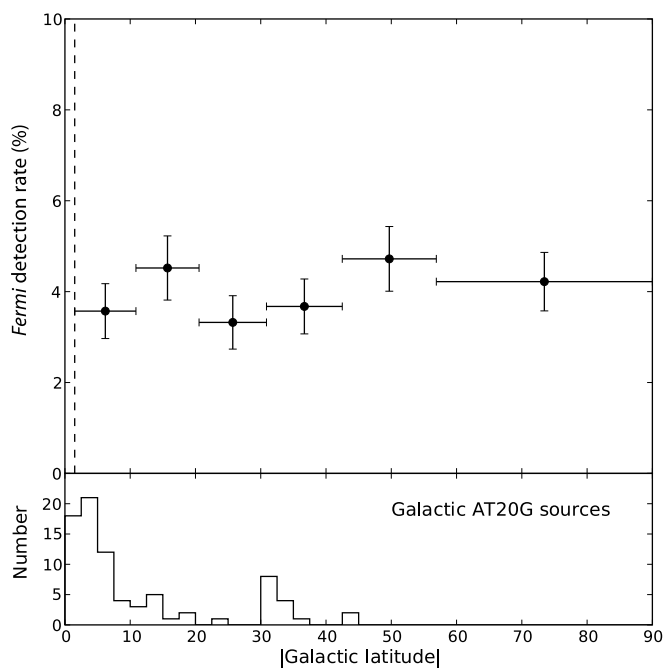

**Figure 4.** Percentage of AT20G sources that have *Fermi* counterparts binned in equal area bins of Galactic latitude. The region close to the plane ($|b| < 1°.5$) is excluded here as that area was not covered in the AT20G survey. Once again the vertical error bars are the counting errors, while the horizontal error bars show the size of each bin. The lower plot shows the distribution of sources flagged as Galactic in the AT20G catalog. These sources have been removed when calculating the detection rate. The dashed line at $|b| = 1°.5$ indicates the limit of the AT20G survey, and lower latitudes have been excluded from this analysis.

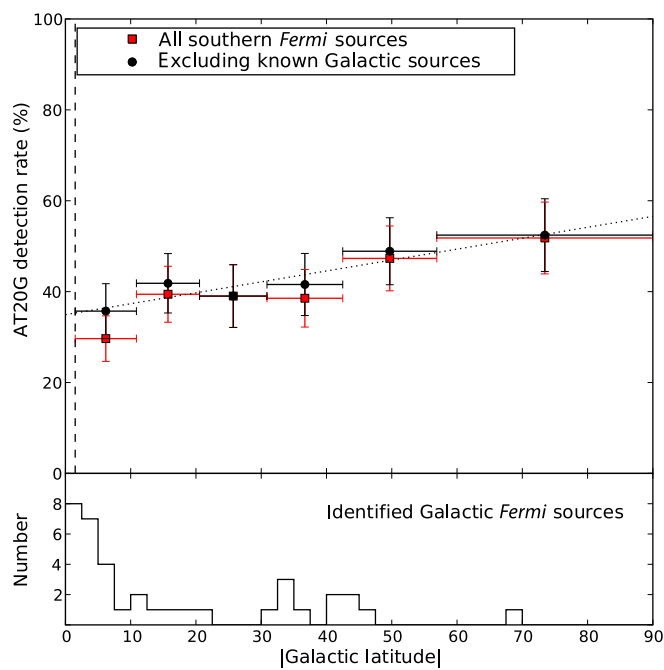

**Figure 5.** AT20G detection rate binned in Galactic latitude. The black circles show the detection rate of AT20G sources in the 1FGL-South data set (where known Galactic sources have been removed), while the red squares show the detection rate when these sources are included. The lower plot shows the histogram of sources defined as Galactic in the 1FGL catalog that have been added when plotting the red squares. The dotted line shows the linear fit to the black points.

(A color version of this figure is available in the online journal.)

The *Fermi* detection rate of ∼4% is roughly constant with Galactic latitude, as would be expected if the AT20G sources being matched are predominantly AGNs. A linear regression analysis reveals a gradient of 0.009, and we cannot rule out the null hypothesis that the slope is flat. The flat distribution provides further evidence that the AT20G–*Fermi* associations are genuine matches. If there were many spurious identifications, we would expect to see a peak close to the Galactic plane, where the surface density of *Fermi* sources is higher than at the Galactic poles.

We can now turn this around to look at the AT20G detection rate of sources in the 1FGL-South data set (Figure 5), after first removing the 37 1FGL sources that are associated with Galactic objects such as pulsars and pulsar wind nebulae (none of these 37 Galactic *Fermi* sources are detected in the AT20G survey). The AT20G detection rate of 1FGL sources is *not* constant with Galactic latitude, but decreases at lower Galactic latitudes. Here, a linear regression fit has a slope of 0.23, and we can rule out that the slope is flat at the 99% confidence level. This strongly suggests that the 1FGL catalog contains a mixture of extragalactic and (unidentified) Galactic sources, whose relative numbers change with $|b|$. The red squares in Figure 5 show the detection rate when the known Galactic sources are added back to the 1FGL-South sample.

To calculate the minimum number of Galactic sources still unidentified, we assumed that there were no unidentified Galactic sources toward the Galactic poles. Taking the detection rate of 52% given in the last bin in Figure 5 to be the expected AT20G detection rate across the whole sky means that there are at least

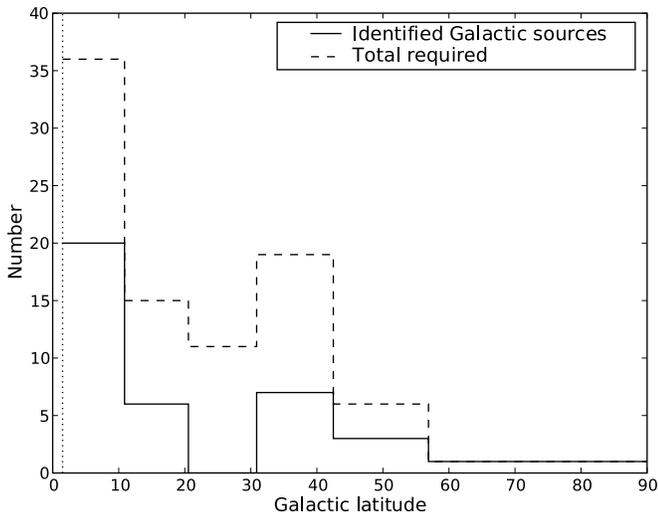

**Figure 6.** Minimum number of Galactic *Fermi* sources needed to make the AT20G detection rate constant with the Galactic latitude.

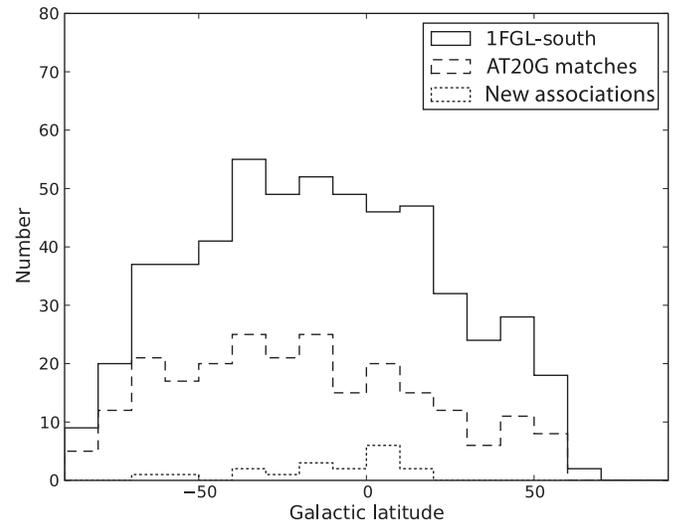

**Figure 7.** Galactic latitude distribution of all 1FGL-South sources (solid line), sources with AT20G matches (dashed line), and sources that have been newly identified as counterparts (dotted line).

50 high-latitude ($|b| > 1°.5$) Galactic gamma-ray sources in the southern hemisphere that still remain to be identified. The estimated distribution of these sources is shown in Figure 6. Here, the total number of sources in each bin is the amount needed to make the AT20G detection rate constant with Galactic latitude (Figure 5). The Galactic sources already identified are shown by the solid line.

We performed a two-dimensional (2D) Kolmogorov–Smirnov test comparing the 1FGL-South Galactic latitude distribution with the distribution of *Fermi*–AT20G matches (the solid and dashed lines shown in Figure 7). This showed that they were not drawn from the same population at the 99% significance level. As the AT20G matches are all AGNs, this provides further evidence that not all the Galactic sources in the 1FGL catalog have been identified.

## 4. RESULTS

### 4.1. Flux Comparison

To investigate any correlations between the radio and gamma-ray properties, we plotted the 20 GHz flux against the gamma-ray flux density as shown in Figure 8, revealing a trend of increasing gamma-ray flux density with increasing 20 GHz flux. This was also hinted at by the *Fermi*-LAT Bright Source Catalogue (Abdo et al. 2009b), but the fainter gamma-ray sources were needed to confirm that the trend did exist. Some additional scatter may be introduced due to the fact that the AT20G observations were taken up to 2–5 years previous to the *Fermi* observations. However, it is still notable that there are no bright 20 GHz sources with weak gamma-ray fluxes. The upper limits shown on the left side of the plot represent the 1FGL-South sources that do not have an AT20G counterpart. As discussed in Section 3.5, some of these may be unidentified Galactic sources for which we would not expect to see any correlation with 20 GHz flux.

Spearman-rank correlation tests reveal correlation coefficients of $\rho = 0.44$ and $\rho = 0.46$ for the variable and non-variable sources, respectively. The probabilities of finding these coefficients by chance are $P = 9.0 \times 10^{-6}$ for the variable sources and $P = 1.5 \times 10^{-8}$ for the non-variable sources. For all the *Fermi*–AT20G sources, the correlation coefficient is $\rho = 0.49$ with a chance probability of $P = 7.5 \times 10^{-16}$. These

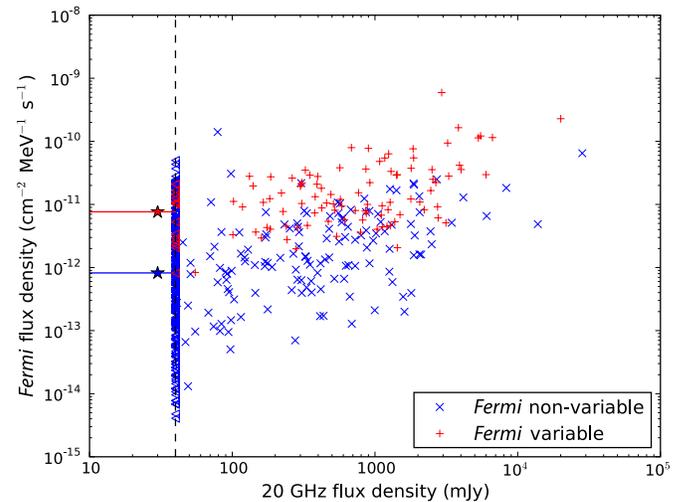

**Figure 8.** 20 GHz flux against *Fermi* flux density for sources classified as variable in 1FGL (red "+") and those not classified as variable (blue "×"). There is a clear trend for the non-variable sources. The *Fermi* sources not detected in AT20G are shown as triangles at the 40 mJy flux limit of the survey (also shown by the dashed line). The median gamma-ray fluxes for the upper limits are denoted by the stars, but it should be noted that these upper limits include a mixture of populations and are not limited to AGN.

(A color version of this figure is available in the online journal.)

numbers are similar to the correlation coefficient of $\rho = 0.42$ with chance probability $P = 4.5 \times 10^{-6}$ between the 8.4 GHz and peak gamma-ray flux given in Giroletti et al. (2010) and Abdo et al. (2009a). The above references also note that a number of biases and selection effects need to be taken into account to test the significance of these correlations. One such example is the bias introduced due to the selection of flat spectrum sources from lower frequency catalogs taken at different epochs, as is done for the CRATES catalog. Using the AT20G survey to find *Fermi* counterparts provides a clean and unbiased sample as it is a blind survey at 20 GHz. Due to the large number of upper limits, a survival analysis approach is needed to investigate the significance of these correlations. However, since many of the upper limits are likely to be Galactic sources that would contaminate any correlation, we have not attempted that in this paper.

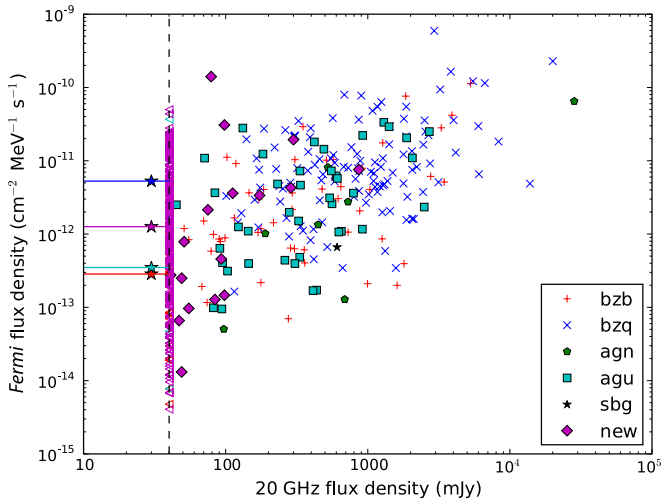

**Figure 9.** 20 GHz flux against, the *Fermi* flux density for the different classes of sources given in the 1FGL catalog. The stars on the left once again show the median gamma-ray fluxes for the *Fermi* sources not detected in AT20G.
(A color version of this figure is available in the online journal.)

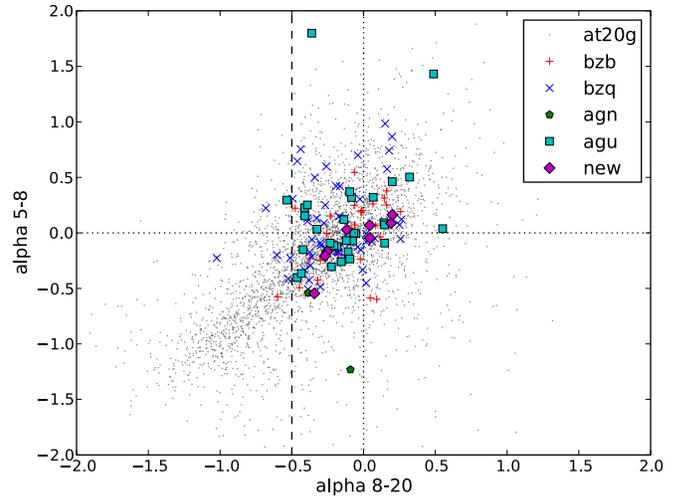

**Figure 10.** Comparison of spectral indices between 4.8 to 8.4 GHz and 8.4 to 20 GHz. The majority of *Fermi* sources are clustered toward the center of the plot implying that they have flat spectral indices from 4.8 to 20 GHz. The gray points show the distribution of all AT20G sources with simultaneous measurements at the three frequencies. The tail of AT20G sources in the lower left corner is the steep spectrum radio sources. The dashed line marks the division of flat and steep spectrum sources.
(A color version of this figure is available in the online journal.)

Figure 8 also shows that the variable *Fermi* sources have higher gamma-ray fluxes than the non-variable sources. This can account for the higher AT20G detection rate for the variable sources mentioned in Section 2, as the correlation between gamma-ray and 20 GHz flux implies that a larger fraction of the brighter gamma-ray sources will be above the 40 mJy flux limit of the AT20G survey.

Figure 9 shows that this trend splits into different population classes. These are classified in the 1FGL catalog (Abdo et al. 2010a) as blazars (either BL Lac objects (bzb) or flat spectrum radio quasars (bzq)), non-blazar active galaxies (agn), active galaxies of unknown type (agu), or starburst galaxies (sbg). The new AT20G associations (those listed in Table 3) are shown by the purple diamonds. Optical spectroscopy is needed to properly classify these objects. Spearman-rank correlation tests give a correlation coefficient of $\rho = 0.31$ and a corresponding chance probability of $P = 0.0006$ for the FSRQ and $\rho = 0.43$, $P = 0.002$ for the BL Lac objects. In comparison, Abdo et al. (2009a) reported correlation coefficients of $\rho = 0.19$, $P = 0.080$ and $\rho = 0.49$, $P = 0.00055$ for the FSRQ and BL Lac objects, respectively. This work used the 8.4 GHz flux and the peak gamma-ray flux above 100 MeV for sources in the *Fermi*-LAT Bright AGN Sample (LBAS).

These results indicate that the basic emission mechanism for the radio and gamma-rays is coupled, with both related to the relativistic jets of the AGN. The radio emission results from synchrotron emission by relativistic electrons while the gamma-ray emission is thought to be created by inverse Compton scattering of photons by similar, or possibly the same, high energy particles (Maraschi et al. 1992; Ghisellini & Madau 1996; Bloom & Marscher 1996). A similar correlation between the gamma-ray and high-radio frequency fluxes was found in Kovalev et al. (2009), who compared the average 100 MeV–1 GeV photon flux with quasi-simultaneous Very Long Baseline Interferometry (VLBI) 15 GHz flux measurements for objects in the *Fermi*-LAT Bright Source Catalogue. There is also the possibility that the Doppler boosting effects are similar in both the radio and gamma-ray regimes. This could account for the strength of the correlation seen in Figures 8 and 9 given that the observations were not simultaneous.

### 4.2. Radio Spectral Indices

The highly variable nature of flat spectrum radio sources makes it difficult to study the spectral index properties without simultaneous data. The AT20G survey is therefore very suitable for investigating these properties since many objects have near simultaneous measurements at 20, 8.4, and 4.8 GHz (separated by approximately one month). Figure 10 shows the comparison of radio spectral indices from 4.8 to 8.4 GHz and from 8.4 to 20 GHz for those objects with simultaneous AT20G data. Comparing the *Fermi* detections with the entire AT20G catalog (shown in gray in Figure 10) highlights the fact that the gamma-ray AT20G sources are generally confined to the flat spectrum population, typical of FSRQ and BL Lac objects.

We have also cross-matched with lower radio frequency catalogs using either the 1.4 GHz NVSS (Condon et al. 1998) or the 843 MHz SUMSS (Mauch et al. 2003) catalogs.[7] These two surveys are ideal, since when they are combined, they cover the entire southern sky at a similar resolution (45 arcsec) to a similar flux limit (approximately 2 and 6 mJy, respectively). Figure 11 shows the distribution in spectral index for a range of different frequencies. As expected the vast majority of sources have flat spectra for all frequency ranges. The median spectral indices for each distribution are $\alpha_8^{20} = -0.16$, $\alpha_5^8 = -0.01$, and $\alpha_1^5 = 0.00$, where $S \propto \nu^\alpha$.

### 4.3. Optical Properties

The accuracy of the AT20G positions allows us to search for optical counterparts in a straightforward manner. The Super-COSMOS Science Archive (Hambly et al. 2001) was searched for the closest source within a 2.5 arcsec radius down to the completeness limit of $b_J = 22$ mag. The Galactic plane ($|b| < 10°$) was excluded in the analysis due to the high density of foreground stars and Galactic dust extinction. There were 35 *Fermi*–AT20G matches excluded as a result. Using this simple approach introduces selection effects, meaning that some identifications may be missed. For example, very nearby galaxies (i.e., Centaurus A, NGC 253, and NGC 4945) are not found because the search radius is too small compared to the angular size of the galaxy. High redshift quasars may also be missed due to the Ly$\alpha$ forest moving into the $b_J$ band and therefore falling below the magnitude limit. A more rigorous method of finding optical counterparts for AT20G sources is ongoing and is discussed in E. K. Mahony et al. (2010, in preparation).

---

[7] The SUMSS catalog excludes $|b| < 10°$ so the second-epoch Molonglo Galactic Plane Survey (MGPS-2; Murphy et al. 2007) was also searched in order to have complete sky coverage. For simplicity, we refer to both of these catalogs combined as the SUMSS catalog.

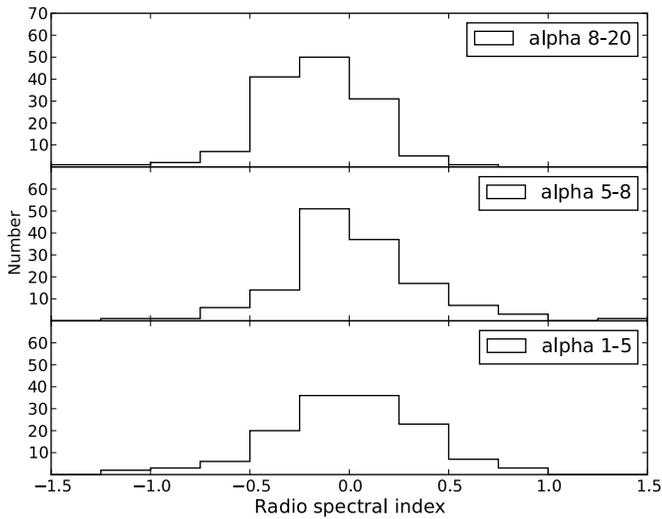

**Figure 11.** Radio spectral index distributions between different frequencies where $S \propto \nu^\alpha$.

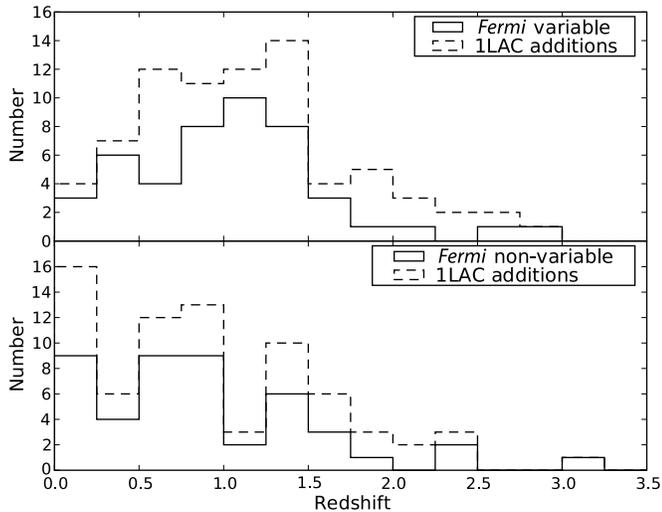

**Figure 12.** Redshift distribution for variable *Fermi*–AT20G sources (top) and non-variable sources (bottom). The solid line shows the redshifts that were found by searching existing databases, and the dashed line shows the distribution of the 1LAC redshifts.

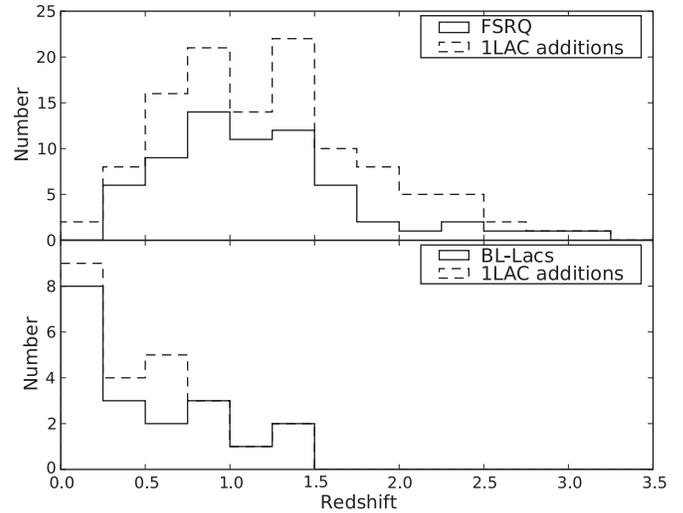

**Figure 13.** Redshift distribution for *Fermi*–AT20G sources. Flat spectrum radio QSOs are shown in the top panel and BL Lac objects in the bottom panel.

Our analysis resulted in 172 (87%) optical counterparts with $b_J \leqslant 22$ mag. Of these, 95 also have redshift information from either the NASA Extragalactic Database,[8] the 6dF Galaxy Survey (Jones et al. 2009), or were obtained from follow-up observations (E. K. Mahony et al. 2010, in preparation). The release of the 1LAC catalog (Abdo et al. 2010b) provided additional redshifts that were not available in existing databases.

---
[8] http://nedwww.ipac.caltech.edu/

These are included in the data table with the reference "1LAC" and are shown by the dashed line in Figures 12 and 13.

Figure 12 shows the distribution of redshifts for both the variable *Fermi* sources and the non-varying sources. Median redshifts are $z = 1.09$ and $z = 0.79$, respectively. Figure 13 shows the redshift distribution for the flat spectrum radio quasars and BL Lac objects. Due to the many unknown biases and selection effects, it is difficult to interpret the significance of these results. For example, many variable sources are classified as flat spectrum radio quasars for which obtaining redshifts at higher redshifts is much easier due to the presence of strong emission lines. It can be more difficult for other extragalactic populations, such as BL Lac objects where the continuum often dominates any features in the optical spectrum. This is shown in Figure 13 where the quasars span a much larger range in redshift with a median of $z = 1.19$ whereas the BL Lac objects have lower redshifts with a median of $z = 0.32$.

These selection biases, along with the incompleteness of the redshift sample, mean that these results are not necessarily representative of the entire *Fermi*–AT20G population.

## 5. CONCLUSIONS

This paper presents results comparing the First *Fermi*-LAT catalog (1FGL) with the AT20G survey. A blind survey at high radio frequencies is an ideal catalog to search for extragalactic gamma-ray counterparts as, like the gamma-ray sky, it is dominated by flat spectrum radio sources (either quasars or BL Lac objects).

Cross-matching the 1FGL catalog with the AT20G catalog resulted in 233 (43%) associations within the 95% *Fermi* confidence ellipse. Monte Carlo tests show that we expect 4.4% to be chance associations, meaning that the final catalog of *Fermi*–AT20G matches is 95% complete and 95% reliable. The detection rate of *Fermi* sources in the AT20G catalog increases strongly with 20 GHz flux density. It ranges from approximately 1% at 100 mJy to almost 100% at 20 GHz fluxes above 10 Jy.

A number of sources were identified that did not have any association information in the published 1FGL catalog. While some of these have been cataloged as "affiliations" in the 1LAC catalog, eight of them are identified here as potential *Fermi* counterparts for the first time. The AT20G source J014137−093001 is also identified as an alternative counterpart to the *Fermi* source 1FGL J0141.7−0929. The remaining AT20G matches are in agreement with the *Fermi* AGN source association.

This confirms that the majority of gamma-ray sources have flat radio spectra as the AT20G survey provides an unbiased sample with no spectral pre-selection. It also verifies that there is no missing gamma-ray population of steep spectrum radio sources.

The distribution of AT20G detections as a function of Galactic latitude reveals that the percentage of AT20G matches decreases toward the Galactic plane. We calculate that there are at least 50 unidentified Galactic gamma-ray sources in the southern hemisphere at latitudes above $|b| = 1°.5$, as the AGN detection rate should not vary with Galactic latitude.

We find a trend of increasing gamma-ray flux density with 20 GHz flux density, which is slightly stronger for the non-variable *Fermi* sources. The majority of the *Fermi*–AT20G matches have flat radio spectral indices across a range of frequencies from 1 GHz to 20 GHz. Of the *Fermi* sources with AT20G detections, 87% also have optical counterparts in SuperCOSMOS. The median redshifts are $z = 1.09$ and $z = 0.79$ for the variable and non-variable *Fermi* sources, respectively. The flat spectrum radio quasars have a median redshift of $z = 1.19$ and the BL Lac objects have a median redshift of $z = 0.32$.

The AT20G survey provides a clean catalog to search for extragalactic gamma-ray sources observed in the southern hemisphere. As such it provides a useful database for identifying gamma-ray counterparts as the *Fermi* mission continues to discover many more of these objects.

This research made use of data provided by the NASA/IPAC Extragalactic Database (NED) which is operated by the Jet Propulsion Laboratory, California Institute of Technology, under contract with the National Aeronautics and Space Administration. It has also made use of data obtained from the SuperCOSMOS Science Archive, prepared and hosted by the Wide Field Astronomy Unit, Institute for Astronomy, University of Edinburgh, which is funded by the UK Science and Technology Facilities Council. The authors thank Andrew Walsh for his helpful advice. We also thank the referee for very useful comments. E.M.S. acknowledges support from the Australian Research Council through the award of an ARC Australian Professorial Fellowship. Partial financial support for this research has been provided to M.M. by the Italian ASI/INAF Agreement I/072/09/0 for the Planck LFI Activity of Phase E2 and I/016/07/0 "COFIS."

*Note Added in Proof.* A recent preprint by Ghirlanda et al. (2010) also cross matches the 1FGL and AT20G catalogues. These authors find a similar (though not identical) list of *Fermi*-AT20G counterparts, but their paper focuses on different aspects of the sample properties and includes SEDs of a few interesting AT20G sources.